\documentclass[a4paper]{article}

\usepackage{INTERSPEECH2022}
\usepackage{xcolor}
\usepackage{cite}

\title{Improving Distortion Robustness of Self-supervised Speech Processing Tasks with Domain Adaptation}

\name{ Kuan Po Huang$^{1\star}$ \qquad Yu-Kuan Fu$^{2\star}$ \qquad Yu Zhang$^{3\star}$ \qquad Hung-yi Lee$^{4\star}$\thanks{$^\star$The authors acknowledge the support of the 2022 Eighth Frederick Jelinek Memorial Summer Workshop.}}
\address{$^{1}$Graduate Institute of Computer Science and Information Engineering, National Taiwan University\\
$^{2}$Department of Physics, National Taiwan University\\
$^3$Google Brain\\
$^{4}$Graduate Institute of Communication Engineering, National Taiwan University
\email{$^{124}$\{f09922005, b07202024, hungyilee\}@ntu.edu.tw, $^3$ngyuzh@google.com}}
\begin{document}

\maketitle
\begin{abstract}
  Speech distortions are a long-standing problem that degrades the performance of supervisely trained speech processing models. It is high time that we enhance the robustness of speech processing models to obtain good performance when encountering speech distortions while not hurting the original performance on clean speech. In this work, we propose to improve the robustness of speech processing models by domain adversarial training (DAT). We conducted experiments based on the SUPERB framework on five different speech processing tasks. In case we do not always have knowledge of the distortion types for speech data, we analyzed the binary-domain and multi-domain settings, where the former treats all distorted speech as one domain, and the latter views different distortions as different domains. In contrast to supervised training methods, we obtained promising results in target domains where speech data is distorted with different distortions including new unseen distortions introduced during testing.
\end{abstract}

\section{Introduction}


Deep learning-based models are powerful in many speech-related applications \cite{yang2021superb}. However, studies have shown that the domain mismatch problem degrades the performance of acoustic models significantly. 
Common domain mismatch scenarios include accented speech, multi-lingual speech, or distorted speech, and this paper focuses on distorted speech.
In the past, there are some rule-based or statistical-based methods for reducing noise \cite{berouti1979enhancement, lim1978all, ephraim1992statistical}. 
There are also some deep learning based approaches \cite{ephraim1995signal, xu2014regression} that try to solve this problem. 
These approaches aim to recover clean speech from noisy speech, but can not guarantee good performance on unseen domain data. 

Self-supervised models, which can be trained with large scale unlabeled data, have become a trend with their ability to output better representations for downstream tasks compared to traditional supervised models. Recently, self-supervised models \cite{ravanelli2020multi, wang2021improving, zhu2022noise} trained with target domain speech proved that they could be robust under domain mismatch scenarios. Performing continual training on self-supervised models \cite{chen2021self, hsu2021robust} with unlabeled target domain data also successfully improved performance on downstream tasks.

Another solution to domain mismatch is training models with domain adversarial training \cite{ganin2016domain} (DAT) based on domain adversarial neural networks (DANN). DANN is a combination of a feature extractor, a downstream model, and a jointly trained domain classifier. The domain classifier distinguishes between the source domain and target domain. With this approach, it is believed that the output representations of the feature extractor can be domain-invariant, so the downstream model can perform comparable results in both source and target domains. Some existing works adopt DAT to solve the domain mismatch problem. For example, automatic speech recognition models \cite{das2021best, na2021accented, sun2018domain, hu2021redat} trained with DAT helped dealt with accented speech. DAT also improved performance for multi-lingual speech emotion recognition \cite{latif2019unsupervised}, text-to-speech synthesis \cite{xin2021cross}, speaker recognition \cite{wang2018unsupervised, huh2020augmentation}, and noise adaptive speech enhancement \cite{liao2018noise}. 

Previous works have shown that DAT is useful in many cases. However, they lack comprehensive studies across multiple self-supervised speech processing tasks. In our work, we combined the concept of DAT and self-supervised training. We proposed to adopt DAT on continually trained self-supervised models to deal with distortion-mismatch scenarios and evaluated the performance on different speech processing tasks of the Speech processing Universal PERformance Benchmark \cite{yang2021superb} (SUPERB). 
Experimental results showed that continually training a self-supervised model on unlabeled distorted data enhances the robustness of models, and DAT can further improve the robustness.
The results showed that performing DAT with a continually trained self-supervised model not only improves performance on speech with distortions seen during training, but also on speech with new types of distortions that did not appear during training.
In some cases, the proposed method even outperformed models trained directly on distorted labeled data.
We also carried out the binary-domain and multi-domain setting with three different domain classifier objectives for DAT, and concluded that in some cases, it is suitable to treat all types of distorted speech as the same domain when we do not have domain knowledge of all speech data. 



\section{Noise-robust Self-supervised Learning}

\subsection{Problem formulation} \label{subsec:problem}

In a typical SUPERB framework, an upstream model serves as a feature extractor $f(s;\theta_f)$.
The input speech waveform $s$ is input to the feature extractor to obtain speech representations $\hat{z}$. 
And a downstream speech processing model serves as a label predictor $y(\hat{z};\theta_y)$ to solve the downstream speech processing tasks.  
Domain mismatch occurs when the testing data has a different distribution compared to the training data. 
A common scenario is that speech distortions are involved during inference. 
This usually causes performance to degrade since the model may not have the ability to generalize to different types of distortions. 
In this work, we aim at adapting models $\theta_f$ to different domains to improve generalizability. 

The problem formulation is stated as follows. Given a source domain dataset $\mathcal{S} = \{ s_1, \cdots, s_N \}$ with $N$ speech utterances and the corresponding ground truth labels $\mathbf{y} = \{y_1, \cdots, y_N\}$, one may train a model supervisedly with the source data and obtain good results on a testing set $\mathcal{C} = \{ c_1, \cdots, c_L \}$ sampled from the same distribution. 
However, if a testing set $\mathcal{C}^\prime = \{ c^\prime_1, \cdots, c^\prime_L \}$ has speech distortions, the performance may suffer from degradation. 
To overcome this domain mismatch problem, assume there is an unlabeled dataset with $K$ types of speech distortions $\mathcal{T} = \{ \{\hat{s}^1_i\}, \{\hat{s}^2_i\} \cdots, \{\hat{s}^K_i\} \}$, where $\{\hat{s}^k_i\}$ is a set of speech utterances with distortions of type $k$. 
In this paper, we considered two scenarios: $\mathcal{C}^\prime \in \mathcal{T}$ and $\mathcal{C}^\prime \notin \mathcal{T}$.
We utilized the unlabeled dataset $\mathcal{T}$ to update the self-supervised model $\theta_f$ in the hope that it can generalize well on the target domain testing set $\mathcal{C}^\prime$. 
This can be performed during the upstream pre-training phase with continual training and during the finetuning phase with DAT. 

\subsection{Stage 1: upstream continual training}
To adapt an upstream model to a new domain, \cite{hsu2021robust, chen2021self} performed continual training by training the pre-trained model with some additional steps on the data of the new domain. 
In our work, we continually train the upstream model with unlabeled data in $\mathcal{T}$ so that the upstream model can gain more model robustness at this stage.

\begin{table*}[ht]
\renewcommand{\arraystretch}{0.8}
\caption{Evaluation results of \textbf{IC}, \textbf{ER}, and \textbf{KS} on the three testing sets of different distortion configurations. The best performances of the given tasks are marked bold.}
\centering
\begin{tabular}{clc|rrr|rrr|rrr}
\toprule
& & & \multicolumn{3}{c|}{\textbf{IC} (Acc)} & \multicolumn{3}{c|}{\textbf{ER} (Acc)} & \multicolumn{3}{c}{\textbf{KS} (Acc)} \\
& & continual & clean & m+g+r & fsd50k & clean & m+g+r & fsd50k & clean & m+g+r & fsd50k \\
\midrule
(a) & baseline  & -  & 99.47  & 96.94 & 97.47 & 63.96 & 57.33  & 60.55  & 97.14  & 93.38 & 93.80  \\
(b) & oracle & - & 99.55 & 99.34 & 98.21 & 70.41 & 69.31 & 69.31 & 97.57 & 96.46 & 95.26\\
\bottomrule
\toprule

(c) & w/o DAT & libri 100hr mgr  & 99.45  & 98.63  & 97.94  & 64.42   & 62.30  & 60.65   & 96.92 & 94.87 & 93.90   \\
(d) & w/o DAT & libri 960hr mgr & 99.39 & 98.84 & 97.89 & 67.28 & 67.47 & 65.62 & 97.12 & 96.11 & 94.77\\
\bottomrule
\toprule
& &  & \multicolumn{9}{c}{Cross Entropy Loss (CE)} \\
(e) & DAT $\lambda=0.01$   & -  & 99.47   & 98.68 & 97.47   & 68.85   & 63.59   & 63.50   & \textbf{97.44}    & 95.26  & 94.64  \\
(f) & DAT $\lambda=0.001$ & -   & 99.60   & 98.60  & 97.63  & 69.10   & 65.90   & 64.29  & 97.24  & 95.65   & 94.55 \\

\bottomrule
\toprule

& &  & \multicolumn{9}{c}{Cross Entropy Loss (CE)} \\
(g) & DAT $\lambda=0.001$ & libri 100hr mgr  & 99.66  & \textbf{99.45}  & \textbf{98.55} & 69.95  & 66.64  & 67.47 & 96.85  & 95.42  & 94.09  \\
(h) & DAT $\lambda=0.001$ & libri 960hr mgr & 99.55 &	99.39 & 98.31 &	\textbf{71.71}	& \textbf{69.12} &	\textbf{69.40} & 97.05 & \textbf{96.27} & \textbf{96.46}\\

\bottomrule
\toprule

& & & \multicolumn{9}{c}{Entropy Loss (E)}  \\
(i) & DAT $\lambda=0.01$   & - & 99.55   & 98.47  & 97.36  & 63.87  & 59.91   & 59.26   & 96.92 & 94.94   & 94.06\\
(j) & DAT $\lambda=0.001$ & -   & 99.58 & 98.27 & 97.52  & 64.15  & 61.75   & 59.54   & 97.05  & 94.87  & 94.13 \\

\midrule
& &  & \multicolumn{9}{c}{Binary Cross Entropy Loss (BCE)}   \\
(k) & DAT $\lambda=0.01$   & -& \textbf{99.68} & 98.39& 97.57 & 66.18 & 64.33 & 62.86  & 96.98& 95.10  & 93.93 \\
(l) & DAT $\lambda=0.001$ & -& 99.60  & 98.97 & 97.89 & 68.76 & 64.52 & 64.52  & 97.27 & 95.59 & 93.90  \\

\bottomrule
\end{tabular}
    \label{tab:results1}
\end{table*}


\subsection{Stage 2: domain adversarial training}
To train the SUPERB framework in a domain adversarial manner, a domain classifier $d(\hat{z};\theta_d)$ is introduced to predict the domain of the original speech input $s$. The parameters $\theta_f, \theta_y, \theta_d$ of the feature extractor, label predictor, and domain classifier are optimized according to the following equations:
\begin{equation}
    \theta_y \leftarrow \theta_y - \alpha \frac{\partial L_y}{\partial \theta_y},\qquad \theta_d \leftarrow \theta_d - \beta \frac{\partial L_d}{\partial \theta_d}
\end{equation}
\begin{equation}
    \theta_f \leftarrow \theta_f - \eta (\frac{\partial L_y}{\partial \theta_f} - \lambda\frac{\partial L_d}{\partial \theta_f})
    \label{eq:lambda}
\end{equation}
where $L_f$, $L_y$, and $L_d$ are the losses of the feature extractor, label predictor, and domain classifier, respectively. $\alpha$, $\beta$, and $\eta$ are the learning rates for gradient descent. The feature extractor has a negative term of the domain classifier loss in its objective to ensure it is trained against the domain classifier to produce domain-indistinguishable representations. The gradient reversal parameter $\lambda$ scales this negative loss term, and is a hyper parameter to be searched during training. For the objective of the domain classifier, there are different designs according to the two different settings elaborated in the following sections.


\subsubsection{Binary-domain setting}
Under the binary domain setting, speech data are categorized into only two domains: clean speech in $\mathcal{S}$ and speech with distortions added in $\mathcal{T}$. 
Under this setting, the domain classifier is optimized with the binary cross entropy objective to distinguish between clean and distorted speech.  
\textbf{Binary cross entropy} loss is denoted as $\mathcal{L}_{BCE}$: 
\begin{equation}
    \mathcal{L}_{BCE} = -\frac{1}{N+M}\sum_{i=1}^{N+M} d_i \cdot \log p_i + (1-d_i) \cdot \log (1 - p_i)
\end{equation}
where $p_i$ is the output of the domain classifier after a sigmoid operation, and $d_i$ is the domain class label for the $i$-th speech utterance. $N$ and $M$ are the quantity of the clean speech and the distorted speech examples, respectively. The goal of training adversarially with this objective is to obtain speech representations so that the domain classifier cannot distinguish the clean and noisy speech. 

\subsubsection{Multi-domain setting}
Under the multi-domain setting, speech data are categorized into $K+1$ domains, where the $K$ kinds of distorted speech are viewed as $K$ different target domains, and the clean speech forming the source domain. Two objectives, cross entropy loss and entropy loss, are involved under this setting. \\
\textbf{Cross entropy} loss is denoted as $\mathcal{L}_{CE}$:
\begin{equation}
    \mathcal{L}_{CE} = -\frac{1}{N+M}\sum_{i=1}^{N+M} \sum_{k=0}^K d^k_i \log p^k_i
\end{equation}
$d^k_i$ is the $k$-th entry of the one-hot domain label vector of the $i$-th speech utterance. $p^k_i$ is the $k$-th entry of the output of the domain classifier after softmax. In the process of DAT, maximizing the cross entropy loss results in misclassification of the domain classifier. \\
\textbf{Entropy} loss is denoted as $\mathcal{L}_{E}$ as shown below.
\begin{equation}
    \mathcal{L}_{E} = -\frac{1}{N+M}\sum_{i=1}^{N+M} \sum_{k=0}^K p^k_i \log p^k_i
\end{equation}
By maximizing this objective during DAT, the domain classifier is prone to output uniform probability distributions, in other words, making it output similar probability values for each domain class.



\begin{table*}[ht]
\setlength\tabcolsep{4pt}
\renewcommand{\arraystretch}{0.85}
\caption{Evaluation results of \textbf{SID} and \textbf{ASR} on the three testing sets of different distortion configurations. For \textbf{ASR}, the results on CHiME3 are also reported.}
\centering
\begin{tabular}{clc|rrr|rrrrrrrr}
\toprule
& & & \multicolumn{3}{c|}{} & \multicolumn{8}{c}{\textbf{ASR} (WER)}\\
& &  & \multicolumn{3}{c|}{\textbf{SID} (Acc)} & \multicolumn{2}{c}{clean} & \multicolumn{2}{c}{m+g+r} & \multicolumn{2}{c}{fsd50k} & \multicolumn{2}{c}{CHiME3}\\
& & continual & clean & m+g+r & fsd50k & w/o & w/ LM & w/o & w/ LM & w/o & w/ LM & w/o & w/ LM\\
\midrule
(a) & baseline  & -  & 84.97 & 65.51 &	77.61 &	6.72 & 4.88 & 10.16 & 7.94	 & 9.62 & 7.57 & 33.4 & 29.26\\
(b) & oracle & - & 86.63 & 80.05 & 82.74 & 5.17 & 4.18 & 6.57 & 5.37 & 6.69	& 5.45 & 22.98 & 20.57\\
\bottomrule
\toprule

(c) & w/o DAT  & libri 100hr mgr & 87.02	& 70.91 & 80.96 & 6.23 & 4.87	& 8.04	& 6.47 & 7.90 & 6.38 & 27.82 & 24.27\\
(d) & w/o DAT  & libri 960hr mgr &86.40 & 74.46 & 81.47	& 5.92	& 4.84 &	7.19 & 6.00& 7.15 & 5.87 & 23.83 & 20.81\\
\bottomrule
\toprule

& &  & \multicolumn{11}{c}{Cross Entropy Loss (CE)} \\
(e) & DAT $\lambda=0.01$  & -  & 86.49&	71.82 &	79.76 &	6.16 &	4.60	&11.74&	9.97&	11.39&	12.27&	44.43&	41.51\\
(f) & DAT $\lambda=0.001$ & -  & 87.44 &	72.28 &	79.94 &	6.29&	5.77&	9.70	&9.00 &	9.67&	8.68&	32.98	& 29.03\\

\bottomrule
\toprule

& &  & \multicolumn{11}{c}{Cross Entropy Loss (CE)} \\
(g) & DAT $\lambda=0.001$ & libri 100hr mgr & 88.70 &	79.59 &	83.57 &	5.75 &	4.82 &	7.30 &	6.21 &	7.21 &	6.15 &	25.60 &	22.73\\
(h) & DAT $\lambda=0.001$ & libri 960hr mgr & 89.08	& \textbf{80.27} &	\textbf{85.04} &	\textbf{5.49}	& 4.61 &	\textbf{6.82} &	\textbf{5.69} &	\textbf{6.67} &	\textbf{5.62} &	\textbf{23.44}&	\textbf{20.71}\\

\bottomrule
\toprule

& &  & \multicolumn{11}{c}{Entropy Loss (E)} \\
(i) & DAT $\lambda=0.01$  & - & 90.01 & 75.23 & 84.04 & 5.59 & \textbf{4.45}  & 11.79  & 10.23  & 8.64  & 7.25 & 40.02 &	37.45\\
(j) & DAT $\lambda=0.001$ & - & 90.50 & 73.43 & 84.46 & 6.01 & 4.59  & 9.97& 7.74& 8.88 & 7.31 & 32.77 & 28.90\\

\midrule
& &  & \multicolumn{11}{c}{Binary Cross Entropy Loss (BCE)} \\
(k) & DAT $\lambda=0.01$  & - & 89.25 &	73.71 &	82.12  & 6.23  & 4.68 & 9.68  & 7.56 & 9.18  & 7.38& 31.20	& 28.72\\
(l) & DAT $\lambda=0.001$ & - & \textbf{90.96} & 78.69 & 84.67 & 6.35 & 4.72  & 9.45& 7.34  & 9.11  & 7.38 & 32.25 & 27.88\\
\bottomrule

\end{tabular}
 \label{tab:results2}
\end{table*}

\section{Experiment setups}

\subsection{Downstream speech processing tasks and models}
In this work, we conducted experiments on five downstream speech processing tasks in the SUPERB benchmark~\cite{yang2021superb}. 
\textbf{Intent Classification, IC} classifies the utterances into different intent classes. We use Fluent Command Speech \cite{lugosch2019speech} as our dataset, which has three intent classes: action, object, and location. The downstream model of this task is composed of a meanpooling operation followed by a linear layer. 
\textbf{Emotion Recognition, ER} predicts the speaker emotion from speech. We use IEMOCAP \cite{busso2008iemocap} as our dataset, which is composed of 151 videos and 9 emotions. The downstream model of this task is three layers of convolutional networks, one layer of self-attention, and two linear output layers. 
\textbf{Keyword Spotting, KS} detects some predefined keywords in the utterances. The Speech Commands v1.0 \cite{warden2017speech} is used as the dataset. The downstream model is composed of three linear layers. 
\textbf{Speaker Identification, SID} recognizes the speakers of the utterances. VoxCeleb1 \cite{nagrani2017voxceleb} is used as the dataset. The downstream model follows the encoder-decoder architecture, where the encoder is a two-layer transformer followed by self-attention pooling, and the decoder being one linear layer. 
\textbf{Automatic Speech Recognition, ASR} transcribes the given utterances to text. LibriSpeech-clean-100 \cite{panayotov2015librispeech} is used as the training data, and LibriSpeech-dev-clean and LibriSpeech-test-clean are the validation set and testing set. The model architecture is two layers of bidirectional LSTM \cite{hochreiter1997long} with CTC\cite{graves2006connectionist} as the decoding strategy. 

\subsection{Data configurations}
We equally split the original training data of the downstream speech processing tasks into two splits, the clean split and the noisy split. 
We assume only the labels of the clean splits are available for training the downstream models. 
Because the labeled training data used in this paper are only half of the labeled data used in the original SUPERB paper~\cite{yang2021superb}, and the upstream models are not fixed during training, the performances in this paper are not comparable with the previous work.
The noisy splits are distorted to serve as an unlabeled noisy dataset $\mathcal{T}$ in Section~\ref{subsec:problem}. Each speech utterance in the noisy splits contains one of the three distortion types, Musan\cite{snyder2015musan} noise (m), Gaussian noise (g), and reverberation (r). The Musan noise corpus consists of various kinds of real-world noises, such as technical noise, domestic noise, traffic noise, etc. Gaussian noise is a statistical noise with a normal distribution as its probability density function. Reverberation is the phenomenon where sound or energy is preserved in a room or space due to reflection from obstacles. In the target domain split, among all the speech utterances, the three kinds of distortions, Musan noise, Gaussian noise, and reverberation, have the proportions 0.3, 0.4, and 0.3, respectively. Each utterance is only distorted with one kind of distortion. The signal-to-noise ratio (SNR) for additive noise is randomly sampled between 10 and 20 dB. 

In this work, there are three testing set configurations. 
The first configuration is the original testing set for each downstream task provided by the SUPERB benchmark. 
The second configuration is constructed by adding the three kinds of distortions to the speech utterances in the testing split in the same way as $\mathcal{T}$. Thus, in the second configuration, the distortions are seen during training. 
The third configuration is meant to construct a new distortion at the testing time by adding noise utterances from the FSD50k\cite{fonseca2022fsd50k} dataset. FSD50k is a subset of AudioSet\cite{gemmeke2017audio}. It includes 200 sound classes mainly composed of real-world sound events. 
For \textbf{ASR}, we also tested the models on the CHiME3\cite{barker2015third} dataset, which includes real speech data recorded in 5 different locations.

\subsection{Upstream model}
In this work, we applied the HuBERT base model \cite{hsu2021hubert} pre-trained with 960 hours of Librispeech as the upstream model. For some experiments, we performed 60 epochs of continual training on the already pre-trained HuBERT base model with the LibriSpeech 100 hour or 960 hour split with three kinds of distortions (m, g, and r) added. For the data used for continual training, the proportions of different speech distortions are all $0.25$ for Musan noise, Gaussian noise, reverberation, and clean speech. 

For DAT, since the feature extractor needed to be trained adversarially against the domain classifier, we set the parameters of the upstream model trainable. For simplicity, only the last hidden states of the upstream model were used as the input of the downstream model.

\subsection{Domain classifier}
The domain classifier consists of a mean pooling procedure followed by a linear layer to project the features to the dimension of the number of total domains.

\subsection{Training details}
The Adam\cite{kingma2014adam} optimizer is used for training. For all experiments, the learning rate $\eta$ of the upstream model is set to $1\mathrm{e}{-5}$, since the upstream model is already pre-trained and requires smaller steps of parameter updates. For downstream models, the learning rate $\alpha$ of \textbf{KS} and \textbf{SID} are set to $1\mathrm{e}{-3}$ and $1\mathrm{e}{-1}$, respectively. For \textbf{IC}, \textbf{ER}, and \textbf{ASR}, the learning rate $\alpha$ of the downstream models are set to $1\mathrm{e}{-4}$. For the domain classifier, the learning rate $\beta$ is set to $1\mathrm{e}{-4}$. For DAT, we searched the $\lambda$ parameter in Eq. (\ref{eq:lambda}) across $1\mathrm{e}{-1}$ to $1\mathrm{e}{-4}$ in log scale and reported two of the best results for $\lambda = 1\mathrm{e}{-2}$ and $\lambda = 1\mathrm{e}{-3}$.

\section{Results}

Tables \ref{tab:results1} and \ref{tab:results2} are the results of the five downstream speech processing tasks on the testing sets with different distortion configurations. The baselines are trained supervisely with the clean split training data. 
We also trained models with both the clean and noisy split data with the labels of the downstream task given.
Since our proposed approaches do not use the labels of the noisy split, the above setting is considered as an oracle case (denoted as ``oracle'' in the tables).


\subsection{Upstream continual training}
Performing continual training on the pre-trained HuBERT base model with some unlabeled noisy data improves the performances of downstream tasks in most cases (rows (c),(d) vs (a)). 
Continual training on the noises  m, g, and r not only improves the performance on the same distortion configuration but also improves the performance of unseen noise FSD50k and CHiME3 for \textbf{ASR}. 
However, for \textbf{KS}, the performance of the baseline on the clean testing set degrades when the upstream is continually trained. The same goes for the baseline of \textbf{IC}.


\subsection{Domain adversarial training}

Rows (e) and (f) show the results of DAT with cross entropy as the domain classifier objective (without continual training).
By comparing DAT with the baseline results (rows (e)(f) vs (a)), DAT with cross entropy outperforms the corresponding baselines for tasks \textbf{IC}, \textbf{ER}, \textbf{KS} and \textbf{SID} in most cases. 
This suggests that DAT has the ability to solve speech processing tasks with different distortions and can generalize well on speech data with unseen distortions during the testing phase. 
It is also interesting to see that the two methods, DAT and continual training, improve different tasks under different configurations (rows (c)(d) vs (e)(f)).
DAT does not degrade the clean speech performance for \textbf{KS} and  \textbf{IC}. 
This shows that DAT can generalize to speech with distortions while preserving good performance on the clean testing set in most cases.
However, DAT does not improve \textbf{ASR} that much compared to continual training.


Rows (g) and (h) are the results of DAT with continual training.
DAT with continual training outperforms the baseline (rows (g)(h) vs (a)) and individual approaches (rows (g)(h) vs (c)(d)(e)(f)) in most cases.
Note that for all of the speech tasks, there are cases where the performance of DAT with continual training exceeds the oracle in the unseen domain (rows  (g)(h) vs (b)). 
This gives an insight into the potential of integrating continual training and DAT to better handle the domain shift problem when new domains are introduced.



\subsection{Different objective functions}
In this section, we compare the results of DAT using different objectives for training the domain classifier (rows (e)(f)(i)(j)(k)(l)). 
Under the multi-domain setting, for \textbf{ER} and \textbf{KS}, adopting cross entropy as the DAT objective yields the best performance in most of the cases. 
This suggests that for these two tasks, the feature extractor focuses on producing representations that cause the domain classifier to misclassify the domain but does not ensure uniform distributions for the domain classifier output. 
However, this is not the case for the results of \textbf{SID}. Applying DAT with entropy loss outperforms training with the cross entropy objective in every domain. 
This shows that the upstream model adversarially trained to maximize the entropy loss of the domain classifier is more suitable for SID. 
Under the binary-domain setting, DAT with binary cross entropy loss performs well for the \textbf{IC} and \textbf{SID} task. The implications of these results are that it is appropriate to view all kinds of distorted speech as the same domain, meaning that the goal of the upstream model is to output representations so that the domain classifier cannot tell whether the original waveform contains distortions or not. This is useful when there is no knowledge of the distortion types for the speech utterances during training.


\section{Conclusions}
DAT improves the robustness of self-supervised speech processing models. It leverages the unlabeled target domain data to generalize well in different domains. In the future, we will extend the training procedure to a multi-task scenario, where DAT can deal with different tasks and distortions simultaneously. 


\bibliographystyle{IEEEtran}

\bibliography{mybib}


\end{document}